\documentclass[showpacs,preprintnumbers,amsmath,amssymb,nofootinbib]{revtex4}
\usepackage{graphicx}
\usepackage{epsfig}
\usepackage{bm}
\usepackage{amsfonts}

\newcommand{\s}{\sigma}
\newcommand{\eq}{\begin{eqnarray}}
\newcommand{\eqx}{\end{eqnarray}}
\newcommand{\ba}{\begin{equation}}
\newcommand{\ea}{\end{equation}}

\newcommand{\bit}{\begin{itemize}}
\newcommand{\eit}{\end{itemize}}

\newcommand{\fp}{f_\phi}
\newcommand{\fpp}{f_{\phi\phi}}
\newcommand{\fs}{f_\sigma}

\begin{document}

\title{Braneworld models with a non-minimally coupled phantom bulk field: a simple way to obtain the $-1$-crossing at late times}

\author{M. R. Setare}\email{rezakord@ipm.ir}
 \affiliation{Department of Science, Payame Noor University,\\
Bijar, Iran }

\author{E. N. Saridakis } \email{msaridak@phys.uoa.gr}
 \affiliation{Department of Physics,
University of Athens,\\ GR-15771 Athens, Greece}

\begin{abstract}
We investigate general braneworld models, with a non-minimally
coupled phantom bulk field and arbitrary brane and bulk matter
contents. We show that the effective dark energy of the
brane-universe acquires a dynamical nature, as a result of the
non-minimal coupling which provides a mechanism for an indirect
``bulk-brane interaction'' through gravity. For late-time
cosmological evolution and without resorting to special ansatzes
or to specific areas of the parameter space, we show that the
$-1$-crossing of its equation-of-state parameter is general and
can be easily achieved. As an example we provide a simple, but
sufficiently general, approximate analytical solution, that
presents the crossing behavior.
\end{abstract}

 \maketitle

\section{Introduction}

According to cosmological observations our universe is undergoing
an accelerating expansion, and the transition to the accelerated
phase has been realized in the recent cosmological past
\cite{observ}. In order to explain this remarkable behavior, and
despite the intuition that this can be achieved only through a
fundamental theory of nature, we can still propose some paradigms
for its description. Thus, we can either consider theories of
modified gravity \cite{ordishov}, or introduce the concept of dark
energy which provides the acceleration mechanism. The dynamical
nature of dark energy, at least in an effective level, can
originate from various fields, such is a canonical scalar field
(quintessence) \cite{quint}, a phantom field, that is a scalar
field with a negative sign of the kinetic term \cite{phant}, or
the combination of quintessence and phantom in a unified model
named quintom \cite{quintom}. The advantage of this combined model
is that although in quintessence the dark energy equation-of-state
parameter remains always greater than $-1$ and in phantom
cosmology always smaller than $-1$, in quintom scenario it can
cross $-1$.

However, there are strong arguments supporting that the
$-1$-crossing in a single, minimally-coupled scalar-field model,
is unstable under cosmological perturbations realized on
trajectories of zero measure \cite{12}. This feature, together
with additional theoretical evidences such are quantum-correction
incorporation and renormalizability, raised the interest for the
investigation of models where the scalar fields are non-minimally
coupled to gravity \cite{11}.

On the other hand, brane cosmology (motivated by string/M theory),
according which our Universe is a brane embedded in a higher
dimensional spacetime \cite{Rubakov83,RS99}, apart from being
closer to a higher-dimensional fundamental theory of nature, it
has also great phenomenological successes and a large amount of
current research heads towards this direction \cite{branereview}.
The bulk space can contain only a cosmological constant
\cite{Binetruy:1999hy}, but string theoretical arguments led to
the insertion of bulk scalar fields, since at the level of the
low-energy  $5D$ theory it is natural  to expect the appearance of
a dilaton-like scalar field in addition to the Einstein-Hilbert
action \cite{lu}. Therefore, the cosmological evolution on the
brane-universe is a combined effect of both the brane matter
content and the bulk scalar field
\cite{Kanti:2000rd,KKTTZ,tetradisbulk1,BT,Bogdanos:2006dt,Saridakis:2007ns}.
In addition,  being closer to scalar-tensor gravity theories and
to the theoretical evidences described above, one can consider the
bulk scalar field to be non-minimally coupled to the $5D$
curvature scalar \cite{BD}.

In the present work we are interested in investigating $5D$
braneworld models with a non-minimally coupled phantom bulk field.
This scenario combines the advantages of   $4D$   phantom
cosmology, together with the advantages of non-minimally coupled
  $4D$   or brane cosmology. In particular, the basic question is
weather we can acquire an effective cosmological evolution on the
brane-universe, where the dark energy equation-of-state parameter
can cross the phantom divide in the near cosmological past,
without choosing special ansatzes for the various quantities, or
moving in specific, small areas of the parameter space. Since it
is already known that in some special cases, non-minimally
coupling in conventional $4D$  cosmology can lead even single
phantom-field models to experience the $-1$-crossing
\cite{Carvalho:2004ty}, it is interesting to see if this behavior
can be preserved in  (closer to a multi-dimensional theory of
nature) brane cosmology. Indeed, it proves that such models not
only do present the aforementioned behavior, in agreement with
observations, but they do so for a large solution sub-class and
without the need of restricting to a specific area of the
parameter space. In particular, under the low-energy (late-time)
assumptions and up to first order in terms of the non-minimal
coupling parameter, the $-1$-crossing can be simply acquired if
the term $\phi\dot{\phi}$ on the brane-universe (where $\phi$ is
the phantom bulk field) changes sign.

However, we have to make a comment about the quantum behavior of
the examined model. As it is known,  the discussion about the
construction of quantum field theory of phantoms is still open in
the literature. For instance in \cite{Cline:2003gs} the authors
reveal the causality and stability problems and the possible
spontaneous breakdown of the vacuum into phantoms and conventional
particles in four dimensions, arising from the energy negativity
(if ones desires to maintain the unitarity of the theory). On the
other hand, there have also been serious attempts in overcoming
these difficulties and construct a phantom theory consistent with
the basic requirements of quantum field theory
\cite{quantumphantom0}, with the phantom fields arising as an
effective description. Since every warped higher dimensional model
is reduced to an effective 4D one in low energies, the
aforementioned discussion concerns the 5D phantom scenario, too.
The present analysis is just a first approach on the $-1$-crossing
in phantom braneworlds. Definitely, the subject of quantization of
such models is open and needs further investigation.

The plan of the work is as follows: In section \ref{model} we
formulate general braneworld models with a non-minimally coupled
phantom bulk field and arbitrary brane and bulk matter contents.
In section \ref{gensolbr} we provide the general late-time
evolution on the brane-universe and we present the main result of
our investigation, namely the relation between the
brane-universe's effective equation-of-state parameter with the
values of the bulk field and its derivative on the brane. In
section \ref{apprsolu} we construct a simple but quite general
approximate analytical solution, which experiences the
$-1$-crossing. Finally, section \ref{conclusions} is devoted to
summarize our results.

\section{The model}
\label{model}

We consider a general class of single-brane models, in the
presence of a non-minimally coupled phantom bulk field,
characterized by the action:
\begin{eqnarray}
{\cal{S}}=&\int & d^4xdy \sqrt{-G} \left\{ f(\phi){\cal{R}} -
\Lambda +\frac{1}{2}\left(\nabla\phi\right)^2 -V(\phi) -{\cal{L}}
_B^{(m)}\right\} + \int
 d^4x \sqrt{-g} \left\{ -\sigma+{\cal{L}}_b^
{(m)} \right\}.{\label{Action}}
\end{eqnarray}
$G_{MN}$ describes the $5D$ metric while  $g$ denotes the
associated induced metric on the brane, located at $y=0$ without
loss of generality, and as usual $\Lambda$ is the $5D$
cosmological constant and $\s$ is the brane-tension. The term
${\cal{L}} _B^{(m)}$ stands for possible forms of bulk matter
apart from the phantom field, while ${\cal{L}}_b^{(m)}$ accounts
for the matter content of the brane-universe. For the moment we
consider the coupling $f(\phi)$, to the $5D$ Ricci scalar
${\cal{R}}$, to be an arbitrary, general function.

The Einstein's equations arise by variation of the action:
\begin{equation}
f(\phi)\left( {\cal{R}}_{MN} -\frac{1}{2}G_{MN}{\cal{R}} \right) -
\nabla_M\nabla_Nf(\phi) + G_{MN}\nabla^2f(\phi) = \frac{1}{2}T_
{MN},
\end{equation}
where $T_{MN}$ is the total energy-momentum tensor:
\begin{equation}
T_{MN} = T_{MN}^{(\phi)} + T_{MN}^{(B)} + T_{MN}^{(b)} -G_{MN}
\Lambda -G_{\mu\nu}\delta_M^{\mu}\delta_N^{\nu}\,\sigma \delta(y).
\end{equation}
In the expression above, $T_{MN}^{(\phi)}$ accounts for the
phantom bulk field
\begin{equation}
T_{MN}^{(\phi)} =-\nabla_M\phi\nabla_N\phi +G_{MN}\left( \frac{1}
{2}\left(\nabla\phi\right)^2-V(\phi) \right).
\end{equation}
Similarly, $T_{MN}^{(B)}$ stands for the bulk part of the
energy-momentum tensor, while $T_{MN}^{(b)}$ accounts for the
brane-matter content. Finally, varying the action in terms of the
phantom field, we obtain its evolution equation:
\begin{equation}
-\nabla^2\phi -\frac{dV}{d\phi} + {\cal{R}}\frac{df}{d\phi} -\frac
{d\sigma}{d\phi}\delta(y) = 0.
\end{equation}

For the metric we will use the following form \cite{BDL}:
\begin{equation}
ds^2 = -n^2(y,t)dt^2 + a^2(y,t)\gamma_{ij}dx^idx^j + b^2(y,t)dy^2,
\end{equation}
which corresponds to a maximally symmetric, induced
Friedmann-Robertson-Walker geometry on the brane. Thus, the
isometry assumption along three dimensional $\mathbf{x}$-slices,
including the brane, allows as to consider the phantom bulk field
depending only on the fifth coordinate. Furthermore, using square
brackets to denote the jump of any quantity across the brane
($[Q]\equiv Q(y_{+})-Q(y_{-})$), and assuming $S^1/{\mathbb Z}_2$
symmetry across it, we restrict our interest only in the $[0,
+\infty)$ interval, obtaining
\begin{equation}
\label{junction}
 [Q']=2Q'(0^{+}).
\end{equation}
 This relation is going to be used
in the elaboration of the discontinuities of the derivatives of
various quantities at the location of the brane \cite{manoscy}.

In these coordinates, the phantom energy-momentum tensor writes:
\begin{equation}
T_{MN}^{(\phi)} = \left(\begin{array}{ccc}
-\frac{1}{2}\dot{\phi}^2-\frac{n^2}{2b^2}{\phi'}^2+n^2V & 0 &
-\dot
{\phi}\phi'\\
0 & a^2\gamma_{ij}\left[-\frac{1}{2n^2}\dot{\phi}^2+\frac{1}{2b^2}
{\phi'}^2-V\right] & 0\\
-\dot{\phi}\phi' & 0 & -\frac{1}{2}{\phi'}^2-\frac{b^2}{2n^2}\dot
{\phi}^2-b^2V
\end{array}\right),
\end{equation}
where primes and dots denote derivatives with respect to $y$ and
$t$ respectively. For the brane-universe  we assume that it
contains a perfect fluid with equation of state $p=w_m\rho$, where
$\rho$ and $p$ are its energy density and pressure respectively.
Thus, the brane energy-momentum tensor reads:
\begin{equation}
 T_{MN}^{(b)} = \frac{\delta
(y)}{b}\left(\begin{array}{ccc}
\rho n^2 & 0 & 0\\
0 & pa^2\gamma_{ij} & 0\\
0 & 0 & 0
\end{array}\right).
\end{equation}
Similarly, assuming an ideal fluid for the arbitrary forms of bulk
matter, with energy density $\rho_B$ and pressures $P_B$ and
$\overline{P}_B$ (since the pressure on the fifth dimension can be
different), we acquire:
\begin{equation}
T_{MN}^{(B)} = \left(\begin{array}{ccc}
\rho_Bn^2 & 0 & -n^2P_5\\
0 & P_Ba^2\gamma_{ij} & 0\\
-n^2P_5 & 0 & \overline{P}_Bb^2
\end{array}\right).
\end{equation}
Note that we have allowed for an energy-exchange function $P_5$
between the bulk and the brane-universe \cite{BT}. Finally, all
quantities are considered as functions of $y$ and $t$, apart from
the brane-fluid's $\rho$ and $p$ which depend only on time.

In order to focus on the cosmological evolution on the brane we
use the Gaussian normal coordinates $\left(b(y,t)=1\right)$
\cite{tetradisbulk1}. Thus, the non-trivial five-dimensional
Einstein equations consist of three dynamical:
\begin{eqnarray}
3\left\{ {\frac{{a'}} {a}\left( {\frac{{a'}} {a} + \frac{{n'}}
{n}} \right) - \frac{1} {{n^2 }}\left( {\frac{{\dot a}} {a}\left(
{\frac{{\dot a}} {a} - \frac{{\dot n}} {n}} \right) + \frac{{\ddot
a}} {a}} \right) - \frac{k} {{a^2 }}} \right\} f
-\frac{1}{n^2}\left\{ \ddot{f} + \left(3 \frac{\dot{a}}{a}
-\frac{\dot{n}}{n}\right)\dot{f} \right\} +
 \left(3\frac{a'}{a}+\frac{n'}{n}\right)f' =\nonumber\\
 =-\frac{1}{4}{\phi'}^2-
\frac{1}{4n^2}\dot{\phi}^2-\frac{1}{2}V(\phi)-\frac{1}{2}\Lambda
+\frac{1}{2}\overline{P}_B {\label{aaaa}},
\end{eqnarray}
\begin{eqnarray}
a^2 \gamma _{ij} \left\{ {\frac{{a'}} {a}\left( {\frac{{a'}} {a} +
2\frac{{n'}} {n}} \right) + 2\frac{{a''}} {a} + \frac{{n''}} {n}}
\right\} f  +  \frac{{a^2 }} {{n^2 }}\gamma _{ij} \left\{
\frac{\dot{a}} {a}\left( - \frac{\dot {a}} {a} + 2\frac{\dot {n}}
{n} \right) - 2\frac{\ddot {a}} {a} \right\} f - kf\gamma _{ij} +
\gamma_{ij}\left\{ -\frac{a^2}{n^2}\left[ \ddot{f} +
 \left(2\frac{\dot{a}}{a} -\frac{\dot{n}}{n}\right)\dot{f} \right]
 +\right.\nonumber\\
 \left.
 + a^2\left[ f'' + \left(2\frac{a'}{a}
+\frac{n'}{n}\right)f' \right]  \right\}
 =\frac{a^2}{2}\gamma_{ij}\left[-\frac{1}{2n^2}
\dot{\phi}^2+{\phi'}^2-V(\phi)\right]+\frac{a^2}{2}\gamma_{ij}
(P_B-\Lambda)+\frac{a^2}{2}\gamma_{ij}\delta(y)\left(p-\sigma(\phi)\right),\
\ \ \
\end{eqnarray}
\begin{equation}
\ddot{\phi} +
\left(3\frac{\dot{a}}{a}-\frac{\dot{n}}{n}\right)\dot {\phi}
-n^2\left\{\phi''+\left(\frac{n'}{n}+3\frac{a'}{a}\right)\phi'
\right\}- n^2\frac{d V(\phi)}{d\phi}+n^2{\cal{R}}f'-n^2\frac{d
\s(\phi)}{d\phi}\delta(y)=0,{\label{eqph}}
\end{equation}
and two constraint equations:
\begin{eqnarray} 3\left\{ \left(\frac{\dot{a}}
{a}\right)^2 - n^2 \left(  \frac{a''}{a}  +  \left(\frac{a'}{a}
\right)^2 \right) +  k\frac{n^2 }{a^2 }  \right\} f  -n^2 \left\{
f'' + 3\frac{a'}{a}f' \right\} +
 3\frac{\dot{a}}{a}\dot{f}=\ \ \ \ \ \ \ \ \ \ \ \ \ \  \ \ \ \ \ \ \ \ \ \ \ \ \ \ \ \ \ \ \ \ \ \ \nonumber\\
 =-\frac{1}{4}\dot{\phi}^2-\frac{n^2}{4}{\phi'}^2+
\frac{n^2}{2}V(\phi)+ \frac{n^2}{2}\rho_B +
\frac{n^2}{2}\delta(y)\rho + \frac{n^2}{2}\delta(y)\sigma(\phi) +
\frac{n^2}{2}\Lambda,
\end{eqnarray}
\begin{equation}
3\left( {\frac{{n'}} {n}\frac{{\dot a}} {a} - \frac{{\dot a'}}
{a}} \right) f -\dot{f}' + \frac{n'}{n}\dot{f}
=-\frac{1}{2}\dot{\phi}\phi'- \frac{n^2}{2}P_5{\label{eq5}}.
\end{equation}
Note that we have allowed for a $\phi$-dependence of the brane
tension $\s(\phi)$. Finally, as usual the $5D$ Ricci scalar
${\cal{R}}$ is given by
\begin{equation}
{\cal{R}}=3\frac{k}{a^2} + \frac{1}{n^2} \left\{ 6\frac{\ddot
{a}}{a} + 6\left(\frac{\dot{a}}{a}\right)^2 -6\frac{\dot{a}}{a}
\frac{\dot{n}}{n} \right\} -6\frac{a''}{a} -2\frac{n''}{n}
-6\left(\frac{a'}{a}\right)^2-6\frac {a'}{a}\frac{n'}{n}.
\end{equation}

In order to obtain the boundary conditions for the aforementioned
cosmological system, we integrate the $00$ and $ii$ components of
the $5D$ Einstein equations around the brane, making use of
(\ref{junction}). Thus, we result to the following junction
(Israel) conditions (setting also $n(0,t)=1$ without loss of
generality):
\begin{eqnarray}
\label{junc1}
&&-6f\frac{a_0'}{a_0}+2\left.\frac{df}{d\phi}\right|_{0}\phi_0'=\frac{1}{2}(\rho+\sigma)\\
\label{junc2}
&&4f\frac{a_0'}{a_0}+2n_0'f-2\left.\frac{df}{d\phi}\right|_{0}\phi_0'=\frac{1}{2}(p-\sigma)\\
\label{junc3}
&&2\phi_0'+4\left.\frac{df}{d\phi}\right|_{0}\left(n_0'+3\frac{a_0'}{a_0}\right)=-\left.\frac{d\sigma}{d\phi}\right|_{0},
\end{eqnarray}
where the index $0$ denotes the values of the corresponding
quantities at the location of the brane. For notation
simplification in the following we omit it, i.e $a$, $n$ and
$\phi$ and their derivatives, stand for the corresponding
quantities on the brane. Furthermore, we call $\fp$ and and $\fs$
the terms $\left.\frac{df}{d\phi}\right|_{0}$ and
$\left.\frac{d\sigma}{d\phi}\right|_{0}$ respectively.

Finally, equations (\ref{junc1}) can be re-written as:
\begin{equation}
\frac{a'}{a}=-\frac{1}{2}\frac{(\sigma+2\fp\fs)}{u} + \frac{1}{8}
\frac{(3p-\rho)}{u}-\frac{1}{16}\frac{(\rho+p)}{f} , {\label{Delta
a}}
\end{equation}
\begin{equation}
n'=-\frac{1}{2}\frac{(\sigma+2\fp\fs)}{u} + \frac{1}{8}\frac{(3p-
\rho)}{u} + \frac{3}{16}\frac{(\rho+p)}{f} ,{\label{Delta n}}
\end{equation}
\begin{equation}
\phi'=-\fp\frac{(3p-\rho)}{u}-\frac{(3f\fs-4\sigma \fp)}{u} ,
{\label{Delta phi}}
\end{equation}
where we have defined
\begin{equation}
u\equiv 6f + 16\fp^2.{\label{U}}
\end{equation}

\section{General late-time Cosmological Evolution on the Brane-universe}
\label{gensolbr}

In this section we  develop the formalism for the late-time
brane-evolution investigation, following
\cite{KKTTZ,Bogdanos:2006dt}. First of all, using the boundary
conditions (\ref{junc1})-(\ref{junc3}), together with
({\ref{eq5}}) at $y=0$, we acquire the energy conservation
equation on the brane:
\begin{equation}
\dot{\rho}+3\frac{\dot{a}}{a}(\rho+p)+2P_5=0 {\label{eqrho8}},
\end{equation}
which in the case $P_5\neq0$ describes the direct bulk-brane
energy flow, which can have both signs. Similarly, using equation
({\ref{aaaa}}) at $y=0$, we obtain the cosmological evolution  on
the brane:
\begin{eqnarray}
3f\left[
\frac{\ddot{a}}{a}+\left(\frac{\dot{a}}{a}\right)^2+\frac{k}
{a^2}\right] + \ddot{f}+
3\frac{\dot{a}}{a}\dot{f}-\frac{1}{4}(\dot{\phi})^2-\frac{1}{2}\left
(V+\Lambda\right)=\frac{1}{64u}\left(3p-\rho-4\sigma\right)^2
-\frac{6}{(16)^2}\frac{(\rho+p)^2}{f}-\nonumber\\
-\frac{1}{2}\overline{P}_B +\frac{\fp\fs}{4u}(3p-\rho-4\sigma)
-\frac{3f}{8u}(\fs)^2.
 {\label{fr2bf}}
\end{eqnarray}
Finally, the use of ({\ref{eqph}}) at $y=0$ provides the phantom
field evolution on the brane:
\begin{equation}
-\ddot{\phi}-3\left(\frac{\dot{a}}{a}\right)\dot{\phi}+\frac{dV}{d
\phi}-6\fp\left[\frac{k} {2a^2}+\frac{\ddot{a}}{a}+
\left(\frac{\dot{a}}{a}\right)^2\right]+\phi'\left(n'+3\frac{a'}{a}
\right) +
6\fp\left(\frac{a'}{a}\right)\left(\frac{a'}{a}+n'\right)
=-\hat{\phi}''-2\fp\left(3\frac{\hat{a}''}{a}+\hat{n}''\right).
\end{equation}

The terms $\hat{\phi}''(t), \hat{a}''(t), \hat{n}''(t)$ denote the
unknown, ``non-distributional'' parts of the corresponding
derivatives \cite{Bogdanos:2006dt}. Although they could be set to
zero, we consider them to depend on the structure of the bulk.
$\hat{a}''$ and $\hat{n}''$ can be expressed in terms of
$\hat{\phi}''$ and the standard quantities, i.e. $a(t), \phi(t)$,
their time-derivatives and the various matter densities. Assuming
$\fs=0$ and following \cite{Bogdanos:2006dt}, the corresponding
elimination leads to the evolution equations on the brane:
\begin{equation}
\frac{\ddot{a}}{a}+\left(\frac{\dot{a}}{a}\right)^2 +\frac{k}
{a^2}\approx \frac {\sigma^2}{12 fu} + \frac{\sigma}{24
fu}\left(\rho-3p\right) -\frac{1}{6f} \overline{P}_B
+\frac{1}{6f}(V+\Lambda)+\frac{\dot{\phi}^2}{12f}
{{\left(1+4\fpp\right)}} +\frac{\fp}{3f}\left(\ddot{\phi} +
3\frac{\dot{a}}{a}\dot{\phi} \right) ,{\label{eqa}}
\end{equation}
\begin{eqnarray}
&&-\ddot{\phi}-3\frac{\dot{a}}{a}\dot{\phi}+\hat{\phi}''-\frac{u_\phi}{2u}
\dot{\phi}^2 \approx\nonumber\\
 &&\approx\frac{3k} {a^2}\fp
-\frac{6f}{u}\frac{dV}{d \phi}-\frac{2\fp}{u}\left(3P_B+
\overline{P}_B-\rho_B\right)
 + 10\frac{\fp}{u}(V+\Lambda) + {{4\frac{\fp(u+\fp u_\phi)\sigma}{u^3}
\left(\rho-3p\right)}}  +
 {{\frac{8\fp(u+\fp u_\phi)\sigma^2}{u^3}}},{\label{eqaphi}}
\end{eqnarray}
where $u_\phi=\left.\frac{du}{d\phi}\right|_{0}$ and
$\fpp=\left.\frac{d^2f}{d\phi^2}\right|_{0}$. We mention that in
order to simplify the above two equations we have focused on the
late-time  approximation, namely neglecting $\rho^2$-terms
\cite{KKTTZ,Saridakis:2007cy} since late time is equivalent to
low-energy. Finally, note that the phantom evolution equation
still contains the unknown function $\hat{\phi}''$. In order to
provide a specific analytical solution sub-class we have to use an
ansatz for it, and this will be done in the simple example of the
next section.

Equations (\ref{eqrho8}), (\ref{eqa}) and  (\ref{eqaphi}) describe
the cosmological evolution of a brane-universe with arbitrary
curvature, in the case of arbitrary bulk and brane contents, for a
general coupling $f(\phi)$,  in the low-energy approximation. In
order to proceed we have to make a choice for the form of
$f(\phi)$. We choose the well-studied quadratic form:
\begin{equation}
f(\phi) = M_5^3\left( 1-\frac{\xi}{2}\phi^2 \right)
,\label{ffunction}
\end{equation}
where $\xi$ is the coupling parameter \cite{BD}, and $M_5$ the
$5D$ Planck mass. This ansatz is a good approximation to a general
coupling function for small phantom field values. Using
(\ref{ffunction}) we obtain: $\fp=-M_5^3\xi\phi$ and
$u=6M_5^3\left(1-\frac{\xi}{2}\phi^2\right)+16M_5^6\xi^2\phi^2$.

As a next step we proceed to the dark energy field formulation
\cite{KKTTZ}, introducing the dark energy field $\chi(t)$ in a way
that the second-order equation (\ref{eqa}) is replaced by two
first-order ones:
\begin{equation}
\left(\frac{\dot{a}}{a}\right)^2 =-\frac{k} {a^2}+ 2\gamma\rho +
\chi + \lambda
 ,{\label{klplfr}}
\end{equation}
\begin{equation}
\dot{\chi} + 4\frac{\dot{a}}{a}\left\{ \chi + \frac{1}{12f} \left[
\overline{P}_B -
\frac{1}{2}(1-4\xi)\dot{\phi}^2-V+2\xi\phi\left(\ddot{\phi} +
3\frac{\dot{a}}{a}\dot{\phi} \right) \right]  \right\}  = 4\gamma
P_5 -2\dot{\gamma}\rho -\dot{\lambda},{\label{lalala2}}
 \end{equation}
where $\beta, \gamma $ and $\lambda$  are given from
\begin{equation}
\beta\equiv\frac{1}{24fu}
\end{equation}
\begin{equation}
\gamma\equiv \sigma \beta
\end{equation}
\begin{equation}
\lambda \equiv \frac{1}{12f}\left( \Lambda + \frac{\sigma^2}{2u}
\right).{\label{lambda}}
\end{equation}
 A simple verification of the above
formulation can be obtained by differentiating ({\ref{klplfr}}),
substituting in ({\ref{lalala2}}) and using  ({\ref{eqrho8}}). In
this way the initial equation ({\ref{eqa}}) can be recovered.

Under the constructed formulation, equation (\ref{klplfr})
corresponds to the conventional   $4D$  Friedmann equation of the
(brane)-universe, and the function $\lambda$ is just the effective
  $4D$  cosmological constant of the brane. In the minimal coupling
case, i.e when $\xi=0$, we obtain
$\lambda=\frac{1}{12M_5^3}\left(\Lambda+\frac{\sigma^2}{12M_5^3}\right)=const$
(i.e.  $\dot{\lambda}=0$) and thus the universe's dark energy
behaves like a cosmological constant. In addition, one can
fine-tune $\lambda$ to zero \cite{RS99} as:
\begin{equation}
\lambda=\frac{1}{12M_5^3}\left(\Lambda+\frac{\sigma^2}{12M_5^3}\right)=0,
\end{equation}
which corresponds to the familiar Randall-Sundrum fine-tuning and
to the known brane solution with no dark energy.

In the non-minimally coupling case we observe that  $\lambda$ is
not a constant anymore but it has acquired a dynamical nature
$(\dot{\lambda}\neq0)$ due to the dynamical nature of the coupling
$f(\phi(t))$. Therefore, $\lambda$ corresponds to the effective
  $4D$  dark energy of the brane-universe. Surprisingly enough one can
acquire an analytical expression for its equation-of-state
parameter $w_{eff}$, up to second order corrections in terms of
the coupling $\xi$. In particular, differentiating (\ref{lambda}),
using the ansatz (\ref{ffunction}) and keeping only first order
$\xi$-terms we obtain:
\begin{equation}
\dot{\lambda}-\xi\phi\dot{\phi}\lambda=0\ \ \ \ \Rightarrow     \
\ \ \ \ \
\dot{\lambda}+3\frac{\dot{a}}{a}\left(-\frac{1}{3}\xi\phi\dot{\phi}\frac{a}{\dot{a}}\right)\lambda=0.
\end{equation}
This relation is independent of  $\lambda$ and $\xi$
normalizations (since some authors consider $2M_5^3$ instead of
$M_5^3$) and thus general. Therefore, we can straightforwardly
make the identification:
\begin{equation}\label{weff2}
w_{eff}=-1-\frac{1}{3}\xi\phi\dot{\phi}\frac{a}{\dot{a}}\,.
\end{equation}

Relation (\ref{weff2}) is the main result of the present work, and
provides the equation-of-state parameter for the dark energy of a
brane-universe, embedded into a bulk with a non-minimally coupled
phantom field and arbitrary bulk and brane matter contents, up to
first order in terms of the coupling $\xi$. In the
minimally-coupled case ($\xi=0$) we acquire $w_{eff}=-1$, that is
we verify that the dark energy is simply just a cosmological
constant. However, in the non-minimally coupled model, which is
the case of interest of the present work, we observe that the
obtained cosmological behavior is very interesting. In particular,
if $\phi\dot{\phi}$ is negative, then $-1<w_{eff}$, that is the
universe's dark energy behaves like quintessence. It is
interesting to notice that we obtain an effective   $4D$
quintessence behavior, although we have the presence of a phantom
bulk field. On the other hand, if $\phi\dot{\phi}$ is positive,
then $w_{eff}<-1$ and dark energy behaves like a   $4D$  phantom.

The most interesting case is when $\phi\dot{\phi}$ changes sign
during the cosmological evolution. In this scenario $w_{eff}$
crosses the phantom divide. In addition, if initially we have
$\phi\dot{\phi}<0$ and after some particular moment we have
$\phi\dot{\phi}>0$, then the $-1$-crossing can take place from
above to below, that is consistently with observations. We mention
that the crossing can be achieved with the use of only one bulk
field, thus the constructed model is quite economical. This
feature was already known to hold in $4D$ single-field,
non-minimally coupled cosmology, but only in specific cases
\cite{Carvalho:2004ty}. However, the non-trivial fact that it is
maintained in the higher-dimensional, and thus closer to a
fundamental theory of nature, brane cosmology, and without the
need of restricting to specific, small areas of the parameter
space, provides an additional argument for the significance of the
non-minimal coupling  for the description of nature. Finally, note
that the dynamical nature of the $4D$ effective dark energy and
the possible brane-universe acceleration, is obtained
independently from the possible direct energy flow between the
bulk and the brane. In other words, the non-minimal coupling
provides a mechanism for an indirect ``bulk-brane interaction'',
through gravity.

In conclusion, in this section we have formulated the general
late-time (low-energy) cosmological evolution on the
brane-universe. In order to acquire a specific example one can
either solve the aforementioned equations numerically or choose
some simple ansatzes for the scale factor, the bulk field and the
matter terms, and solve the equations analytically. In the next
section, we use some general but simple ansatzes, in order to
present more transparently the cosmological implications of the
constructed model.

\section{Crossing $-1$: a simple approximate solution}
\label{apprsolu}

Let us now present a simple, but sufficiently general,
approximate, analytical cosmological solution. Since we focus our
analysis at late times, it is reasonable to consider the
widely-used power-law ansatzes. Thus, for the bulk field we
assume:
\begin{equation}
\phi(t)=\frac{C_\alpha}{t^\alpha}+\frac{C_\beta}{t^\beta}\label{phi0},
\end{equation}
with $C_\alpha$, $C_\beta$ being constants. Similarly, for the
scale factor we can safely consider the solution:
\begin{equation}
a(t)=C_3t^{\nu_1}\label{a0}.
\end{equation}
Now we have to choose the exchange term $P_5$ that is present in
(\ref{eqrho8}), as well as the bulk pressure $\overline{P}_{B}$
that appears in (\ref{fr2bf}), which are both functions of time,
corresponding to the values of $\overline{P}_B(y,t)$ and
$P_5(y,t)$ on the brane. The energy-momentum conservation
$\nabla_MT^M_{\,\,N}=0$ cannot fully determine $\overline{P}_{B}$
and $P_5$ and a particular model of the bulk matter is required.
Although we could consider the ansatz \cite{BT} (see also
\cite{CGW}):
\begin{equation}
P_5(t)=C_5\left[\frac{\dot{a}(t)}{a(t)}\right]a(t)^{\nu_2},{\label{Ansatz5}}
\end{equation}
for the investigation of this section and for simplification of
the results it is adequate to set $P_5=0$. For the bulk pressure
we assume $\overline{P}_{B}=const$, and the constant can be
absorbed in a re-definition of the bulk potential. This can be
chosen as \cite{Saridakis:2007ns}:
\begin{equation}
V(\phi)=\frac{\mu^2}{2}\phi^2,{\label{AnsatzV}}
\end{equation}
which is a general form, consistent with the brane stabilization
mechanism \cite{Goldberger99}. Finally, for the matter content of
the universe we have already assumed an ideal fluid, with energy
density and pressure connected by:
\begin{equation}
p=w_m\rho{\label{Ansatzwm}}.
\end{equation}
Thus, using    (\ref{Ansatzwm}) we can easily solve (\ref{eqrho8})
obtaining:
\begin{equation}
\rho=\frac{C_\rho}{a^{3(1+w_m)}},{\label{rhosol}}
\end{equation}
with $a(t)$ given by (\ref{a0}).

We mention that the considered ansatzes for  $a(t)$, $P_5$,
$\overline{P}_{B}$ and $V(\phi)$  are not crucial for the obtained
results, and they are chosen just for presentation reasons.  The
only necessary behavior is a $\phi(t)$ with a derivative that
changes sign during the evolution, as we have discussed in the
previous section. The ansatz (\ref{phi0}) can fulfill  this
conditions in a simple way, but any other ansatz with a derivative
sign-flip could  be equivalently used. Finally, if one use an
arbitrary ansatz for $\phi$ without a derivative sign-flip, then
according to (\ref{weff2}) he will acquire a $w_{eff}$ lying
always on the same side of the phantom divide during cosmological
evolution.

 The final step is the
substitution of the aforementioned ansatzes, together with
(\ref{rhosol}), into the cosmological equations (\ref{klplfr}) and
(\ref{lalala2}). In order to simplify the calculations we consider
only the flat brane-universe case ($k=0$), and since we are
investigating the late-time behavior we only keep terms up to
${\cal{O}}(t^{-2\alpha})$ and ${\cal{O}}(t^{-2\beta})$
\cite{Bogdanos:2006dt}. Finally, as stated in section
\ref{gensolbr}, we keep terms up to  ${\cal{O}}(\xi^2)$.
Therefore, within these approximations, the two cosmological
equations become:
\begin{equation}
\frac{\Lambda}{6M_5^3}+\frac{\s^2}{72M_5^6}-\frac{\nu_1(2\nu_1-1)}{t^2}+\frac{(1-3w_m)\,\s\,C_\rho\,
t^{-3(1+w_m)\nu_1}}{144
M_5^6}+{\cal{O}}(t^{-2\alpha})+{\cal{O}}(t^{-2\beta})+{\cal{O}}(\xi^2)=0,{\label{eq1}}
\end{equation}
\begin{eqnarray}
-\hat{\phi}''+t^{-2-\alpha}\left[C_\alpha\,
\alpha(\alpha+1-3\nu_1)-\frac{\s(1-3w_m)}{9M_5^3}\,C_\rho
C_\alpha\,\xi\right]+t^{-2-\beta}\left[C_\beta\,
\beta(\beta+1-3\nu_1)-\frac{\s(1-3w_m)}{9M_5^3}\,C_\rho C_\beta\,\xi\right]-\nonumber\\
-t^{-\alpha}\left[C_\alpha\mu^2+\frac{5}{3}C_\alpha\Lambda\xi+\frac{2C_\alpha\s^2}{9M_5^3}\xi\right]-
t^{-\beta}\left[C_\beta\mu^2+\frac{5}{3}C_\beta\Lambda\xi+\frac{2C_\beta\s^2}{9M_5^3}\xi\right]+
{\cal{O}}(t^{-2\alpha})+{\cal{O}}(t^{-2\beta})+{\cal{O}}(\xi^2)=0.\
\ \ \ \, {\label{eq2}}
\end{eqnarray}
A last assumption concerns the unknown function $\hat{\phi}''$.
Although its specific form is not important, in order to simplify
the calculations and following \cite{Bogdanos:2006dt} we consider:
\begin{equation}
\hat{\phi}''=\frac{C_{\phi\alpha}}{t^{\alpha+2}}+\frac{C_{\phi\beta}}{t^{\beta+2}}
.{\label{hat1}}
\end{equation}

In order for equation (\ref{eq1}) to be satisfied for all $t$ we
require:
\begin{eqnarray}
&&\nu_1=\frac{2}{3(1+w_m)}\label{nu00}\\
&&\Lambda+\frac{\s^2}{12M_5^3}=0\label{Lss}\\
 &&\nu_1(2\nu_1-1)=\frac{\s(1-3w_m)}{144M_5^6}\,C_\rho.
 {\label{fits1}}
\end{eqnarray}
Note that (\ref{Lss}) is the standard Randall-Sundrum fine-tuning
for a vanishing cosmological constant on the brane, in the
minimally-coupling case in the absence of matter. Under these
conditions, the requirement of satisfaction of (\ref{eq2}) at all
times leads to:
\begin{eqnarray}
&& \mu^2=-\frac{\xi\s^2}{12M_5^3}\\
&&C_\alpha\,
\alpha(\alpha+1-3\nu_1)-16M_5^3C_\alpha\,\nu_1(2\nu_1-1)\xi-C_{\phi\alpha}=0\label{alphafit}\\
&&C_\beta\,
\beta(\beta+1-3\nu_1)-16M_5^3C_\beta\,\nu_1(2\nu_1-1)\xi-C_{\phi\beta}=0.
 {\label{betafit}}
\end{eqnarray}
Conditions (\ref{nu00})-(\ref{betafit}) are necessary in order for
the selected ansatzes to form a self-consistent cosmological
solution. The quadratic equations (\ref{alphafit}) and
(\ref{betafit}) lead to:
\begin{eqnarray}
&& \alpha=\frac{3\nu_1-1}{2}\pm\sqrt{\frac{(3\nu_1-1)^2}{4}+\frac{C_{\phi\alpha}}{C_\alpha}+16M_5^3\left[\nu_1(2\nu_1-1)\right]\xi}\label{alphafit2}\\
&&
\beta=\frac{3\nu_1-1}{2}\pm\sqrt{\frac{(3\nu_1-1)^2}{4}+\frac{C_{\phi\beta}}{C_\beta}+16M_5^3\left[\nu_1(2\nu_1-1)\right]\xi}.
 {\label{betafit2}}
\end{eqnarray}
As we can see, if $\nu_1\geq1/2$, that is according to
(\ref{nu00}) if $w_m\leq1/3$, the exponents $\alpha$ and $\beta$
are real for any value of $\xi$. On the other hand for $\nu_1<1/2$
( i.e $w_m>1/3$) the condition $\alpha,\beta\in \mathbb{R}$ leads
to a restriction in $\xi$-values, namely:
\begin{equation}
\xi\leq \text{min}(\xi_1,\xi_2),{\label{xi1}}
\end{equation}
where
\begin{eqnarray}
&& \xi_1=\xi_c\,\frac{\left[(3\nu_1-1)^2+4\frac{C_{\phi\alpha}}{C_\alpha}\right]}{12\nu_1(1-2\nu_1)}\\
&&
 \xi_2=\xi_c\,\frac{\left[(3\nu_1-1)^2+4\frac{C_{\phi\beta}}{C_\beta}\right]}{12\nu_1(1-2\nu_1)}.
\end{eqnarray}
In these expressions we have used the definition
$\xi_c\equiv\frac{3}{16M_5^3}$, namely the conformal value of the
non-minimal coupling parameter $\xi$. Therefore, in the
(non-conventional) case $w_m>1/3$, the value of $\xi$ must be
sufficiently small.

Let us now examine the cosmological behavior of the simple,
approximate, but quite general, model of this section. According
to (\ref{weff2}), and using the ansatzes (\ref{phi0}), (\ref{a0})
we obtain:
\begin{equation}
w_{eff}=-1+\frac{1}{3\nu_1}\left(\frac{C_\alpha}{t^\alpha}+\frac{C_\beta}{t^\beta}\right)
\left(\frac{\alpha C_\alpha}{t^\alpha}+\frac{\beta
C_\beta}{t^\beta}\right)\xi,{\label{weff3}}
\end{equation}
with $\alpha$ and $\beta$ given by (\ref{alphafit2}) and
(\ref{betafit2}). Note that due to the parameter eliminations
through conditions (\ref{nu00})-(\ref{betafit}), one can either
impose a value for $w_m$ and acquire the value of $C_\rho$, or the
opposite. Since the direct observable is $w_m$ we prefer to use
its value as an input, and thus eliminate $C_\rho$ from
$w_{eff}$'s calculation.

As we have already mentioned, for $\xi=0$ dark energy behaves like
a cosmological constant ($w_{eff}=-1=const$), but the non-minimal
coupling gives it a dynamical nature. Assuming $\phi(t)\geq0$
(i.e. choosing $C_\alpha,C_\beta>0$), then if $\dot{\phi}(t)$
changes sign during the cosmological evolution then $w_{eff}$
crosses the phantom divide $-1$, while if $\dot{\phi}(t)$
preserves the same sign, $w_{eff}$ lies always on one side of the
phantom divide. However, one could consider cases where $\phi(t)$
changes sign or where both $\phi(t)$ and $\dot{\phi}(t)$ do.
Finally, as can be seen from (\ref{weff3}) and
(\ref{alphafit2}),(\ref{betafit2}) the decisive role for the
determination of $w_{eff}$ is played by the $\phi$-parameters
($C_\alpha,C_\beta$), $w_m$ and of course $\xi$, with the  rest
parameters being non-relevant (for realistic choices).

 In order to acquire a better
comparison with the observational data, in the following we
provide the evolution of $w_{eff}$ versus the redshift $z$, given
by $(1+z)=a_0/a(t)$ with $a_0$ the present value, where we  fix
$C_3$ in order to acquire $a_0=1$.
\begin{figure}[ht]
\begin{center}
\mbox{\epsfig{figure=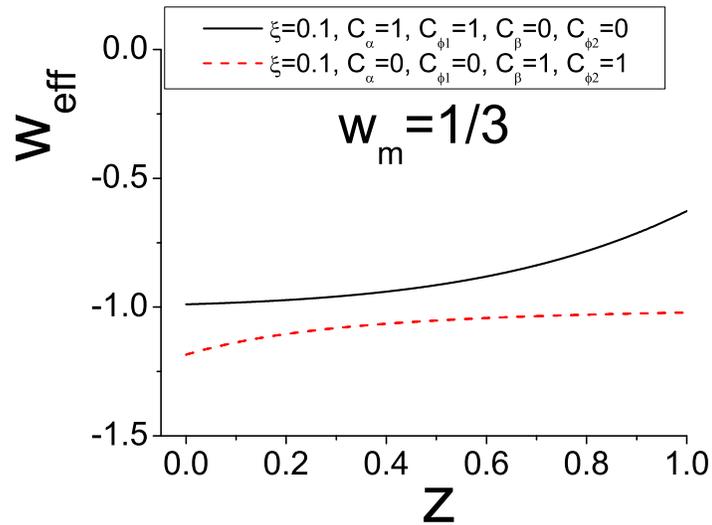,width=9.6cm,angle=0}}
\caption{(Color online) {\it $w_{eff}$ versus $z$, for $w_m=1/3$
and $\xi=0.1$. The upper curve corresponds to $C_\alpha=1,
C_{\phi\alpha}=1, C_\beta=0, C_{\phi\beta}=0$, while the lower one
to  $C_\alpha=0, C_{\phi\alpha}=0, C_\beta=1, C_{\phi\beta}=1$.}}
\label{fig1}
\end{center}
\end{figure}
Furthermore, for the equation-of-state parameter of the matter
content of the universe
 we use the value $w_m=1/3$, since a relativistic matter is the most natural choice.
 This value for $w_m$ has an additional advantage, namely it leads
 to $\nu_1=1/2$ (see (\ref{nu00})) and thus $M_5$ is eliminated
 from the final results (see (\ref{alphafit2}) and
 (\ref{betafit2})). Therefore, since $M_5$ defines the units of
 the present model, its disappearance from the final relations
 ((\ref{nu00}),  (\ref{alphafit2}),
 (\ref{betafit2})) and (\ref{weff3})) allows us to choose the
 units at will. Thus, we choose the units in order for the
 parameters $C_\alpha,
C_{\phi\alpha}, C_\beta, C_{\phi\beta}$ to be of order $1$.

In fig.~\ref{fig1} we depict $w_{eff}$ versus $z$ for $\xi=0.1$,
and two choices of the parameter four-plet $C_\alpha,
C_{\phi\alpha}, C_\beta, C_{\phi\beta}$. This value for $\xi$ is
consistent with  ${\cal{O}}(\xi^2)$-calculations, while the
$C_i$-parameter choices correspond to keeping only one term in
$\phi(t)$, thus its derivative preserves the same sign and
therefore $w_{eff}$ behaves like in $4D$ quintessence (upper
curve) or in
  $4D$  phantom (lower curve) paradigms.

\begin{figure}[ht]
\begin{center}
\mbox{\epsfig{figure=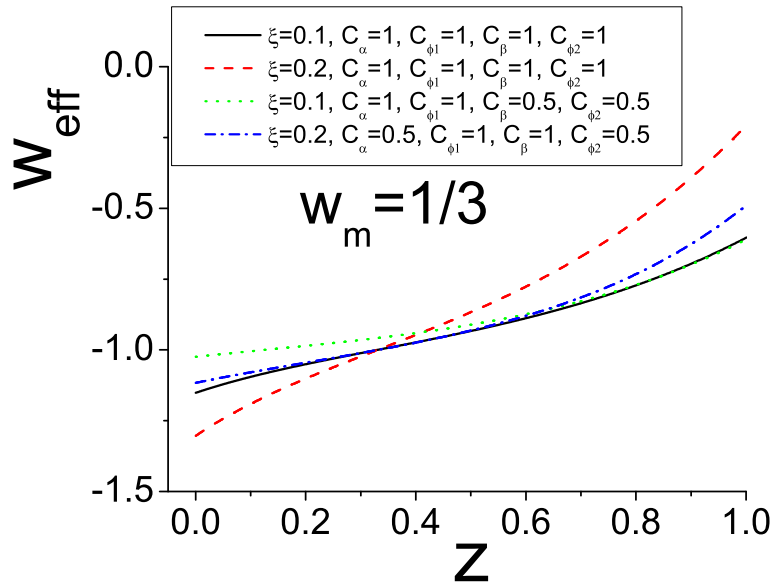,width=10.5cm,angle=0}}
\caption{(Color online) {\it  $w_{eff}$ versus $z$, for $w_m=1/3$.
The values of the parameters $\xi, C_\alpha, C_{\phi\alpha},
C_\beta, C_{\phi\beta}$ are shown in the inset.
 }} \label{fig2}
\end{center}
\end{figure}

In fig.~\ref{fig2} we depict $w_{eff}$ versus $z$ for four choices
of the parameter-group  $\xi, C_\alpha, C_{\phi\alpha}, C_\beta,
C_{\phi\beta}$. In this case we are interested in acquiring a
sign-flip of $\dot{\phi}$, and thus without loss of generality we
keep the plus sign in $\alpha$-solution in (\ref{alphafit2}) and
the minus sign in $\beta$-solution  in (\ref{betafit2}).
As we observe, $w_{eff}$ crosses the phantom divide from above to
below  in the recent cosmological past, as required by
observations. This behavior is qualitatively independent of the
values of the parameters $\xi, C_\alpha, C_{\phi\alpha}, C_\beta,
C_{\phi\beta}$, however the precise values of $z_c$ (redshift at
the crossing) and of $w_{eff0}$ (present value of $w_{eff}$) do
depend on them. Thus, we see that in the simple, but quite
general, example of the present section, the $-1$-crossing can be
achieved relatively easily, without a restriction to a small area
of the parameter space. This feature makes braneworld models with
a non-minimally coupled phantom bulk field a good candidate for
the description of the current universe acceleration.

\section{Conclusions}
\label{conclusions}

In this work we examine general braneworld models, with a
non-minimally coupled phantom bulk field and arbitrary brane and
bulk matter contents. Imposing the low-energy, that is late-time
assumptions, and performing the calculations up to first order in
the non-minimal coupling $\xi$, we provide a general relation that
connects the equation-of-state parameter $w_{eff}$ of the $4D$
effective dark energy, with the values of the phantom field $\phi$
and its derivative at the location of the brane-universe. For
$\xi=0$, $w_{eff}$ is always $-1$ and the dark energy of the
brane-universe behaves like a cosmological constant, as expected.
However, when the non-minimal coupling is switched on, the
brane-universe's dark energy acquires a dynamical nature.

In particular $w_{eff}$ is related to $\xi$, to the scale factor
and its derivative, and to  $\phi$ and its derivative on the brane
(relation (\ref{weff2})). Thus, if $\phi\dot{\phi}$ preserves the
same sign during the cosmological evolution, then $w_{eff}$
remains always on the same side of the phantom divide and the
universe's dark energy behaves like a quintessence or conventional
phantom. On the other hand, if $\phi\dot{\phi}$ experiences a
sign-flip at some particular time, then $w_{eff}$ crosses $-1$, in
agreement with observations. This behavior is general and, up to
first order in $\xi$, it is independent of the bulk and brane
matter contents. That is, the non-minimal coupling provides a
mechanism for an indirect ``bulk-brane interaction'', through
gravity. Furthermore, the crossing behavior appears without the
need of special ansatzes for the various quantities, or of the
restriction to specific, small areas of the parameter space.

In the present model, the $-1$-crossing arises from a single,
non-minimally coupled phantom bulk field, and thus it is more
economical than the multi-field models of conventional cosmology.
This behavior was already known to hold in $4D$ single-field,
non-minimally coupled cosmology, but only for specific cases
\cite{Carvalho:2004ty}. The fact that it is not only maintained in
the higher-dimensional (and thus closer to a fundamental theory of
nature) brane cosmology, but it appears for a much larger solution
sub-class and parameter sub-space, is definitely a novel feature
and an advantage of the model. This result provides an additional
argument for the significance of the non-minimal coupling between
gravity and scalar fields, for the description of nature. However,
the possible quantization difficulties of such phantom scenarios
is an open problem and needs further investigation.\\

\paragraph*{{\bf{Acknowledgements:}}}
One of us (E.N.S) wishes to thank Institut de Physique
Th\'eorique, CEA, for the hospitality during the preparation of
the present work.

\end{document}